\documentstyle[12pt]{article}
\newcommand{\gsim}{\buildrel > \over {_\sim}}
\newcommand{\lsim}{\buildrel < \over {_\sim}}
\newcommand{\ie}{{\it i.e.}}
\newcommand{\eg}{{\it e.g.}}
\newcommand{\cf}{{\it c.f.}}
\newcommand{\etal}{{\it et al.}}
\newcommand{\gev}{{\rm GeV}}
\newcommand{\jpsi}{J/\psi}

\begin{document}

\rightline{\vbox{\halign{&#\hfil\cr
                 &NORDITA - 95/78 P\cr
                 &hep-ph/9511411\cr}}}
\vspace{0.2in}

\begin{center}
{\Large {\bf NUCLEAR EFFECTS IN CHARM PRODUCTION\footnote[2]{Contribution to
the
ELFE Summer School and Workshop, Cambridge, UK, July 1995.}
}}\\
\vspace{.5cm}
{\large Paul Hoyer} \\
\medskip
{\it Nordita, Copenhagen, Denmark } \\

\vspace{1.0cm}

\large {\bf ABSTRACT}

\end{center}

I briefly review our understanding of the nuclear target dependence of charm
and
charmonium production, in view of charm physics at Elfe.


\vspace{1cm}

\setlength{\topmargin}{-1.5cm}
\setlength{\textheight}{25.0cm}
\setlength{\baselineskip}{0.75cm}

\noindent
{\large\bf Charm as a probe of hard interactions}

\noindent
The charm quark mass $m_c \simeq 1.5\,\gev = (\frac{1}{7}\,{\rm fm})^{-1}$ is
large compared to the QCD confinement scale of $\Lambda_{QCD} \simeq .2\,\gev
= 1\,{\rm fm}^{-1}$. Hence charm physics is `hard' physics, as evidenced
already by the very narrow total width (86 keV) of the $\jpsi$. On the other
hand, since the charm mass scale exceeds $\Lambda_{QCD}$ by less than an order
of magnitude, higher order (in $\alpha_s$) and higher twist (in $1/m_c^2$)
corrections can be expected to be sizable. From the point of view of precise
predictions (\eg, of the total charm cross section) that is a disadvantage.
On the other hand, it also implies sizable signals for new physics
beyond the leading twist approximation. By concentrating on relative rates
(such
as the dependence on the atomic number $A$ of a nuclear target), much of the
uncertainty in the absolute prediction is eliminated. Moreover, Nature has
been kind enough to supply us with another readily available heavy quark,
the b quark with mass $m_b \simeq 5\,\gev = (\frac{1}{25}\,{\rm fm})^{-1}$.
Since QCD processes are flavor blind, the quark mass dependence of observables
reveals whether we are dealing with a leading or a higher twist process.

{}From an observational point of view charm has an important advantage
compared to other hard probes such as large $p_\perp$ jets. Jets cannot in
practice be detected below $p_\perp \simeq 5\,\gev$, since their transverse
momentum is distributed over many soft pions. On the contrary, there is no
ambiguity in measuring the momentum of $D$, $D^*$ or $\Lambda_c$ hadrons.
The added bonus of being able to study charmonium states ($\jpsi,\ \psi',\
\chi_c,\ldots$) has no counterpart in jet physics, and has turned out to be
extraordinarily interesting. The standard QCD factorization theorem for hard
processes \cite{qs} is not applicable in a situation where the heavy quarks
are constrained to have low relative momenta, being replaced by a more
sophisticated expansion in powers of their relative velocity \cite{bbl}. The
tentative models of charmonium production (Color Evaporation, Color
Singlet, Color
Octet,...) are in fact having a hard time describing the data \cite{gas}.

A prime purpose of physics at Elfe will be to investigate the space-time
dynamics
of hard collisions. Through the use of nuclear targets, additional information
about the short time development of the scattering process can be obtained. The
moderate hardness scale offered by the charm quark implies that charm physics
will be an important part of this endavour. In the following I shall touch upon
some of the topical questions of charm physics which are related to nuclear
target dependence.

\vspace{.5cm}
\noindent
{\large\bf Nuclear target dependence}

\noindent
At leading twist, charm production proceeds via subprocesses such as
$\gamma^*g \to c\bar c$ in leptoproduction, and $gg \to c\bar c$ in
hadron collisions. The measured total charm cross section is found to be in
rough
agreement with QCD expectations, within the rather large uncertainties due to
scale dependence and higher order perturbative corrections \cite{gas,fmnr}.

The hadroproduction of $D$ mesons on nuclear
targets has been well measured in proton collisions at 800 GeV.
Parametrizing the $A$-dependence as $\sigma(pA \to D+X) \propto
A^\alpha$, E769 \cite{e769} finds an average $\alpha = 1.00 \pm .05 \pm .02$
for
$0<x_F(D)<0.8$. The E789 Collaboration \cite{e789} similarly obtains $\alpha =
1.02 \pm .03 \pm .02$, for an average $< x_F > = 0.031$. Thus charm production
(in the region of small $x_F$ which dominates the total cross section) is
consistent with the leading twist expectation of being additive on all
partons in
the nucleus. This is to be compared to the typical values $\alpha = 0.7 \ldots
0.8$ found for light hadrons ($\pi,\ K,\ p \ldots$) in this kinematic region
\cite{dsb}. Clearly charm production qualifies as a hard process.

The $A$-dependence of inclusive charmonium $(\jpsi,\ \psi')$ and bottomonium
production \cite{cha} reveals that the reaction dynamics is considerably more
complicated than for open charm.

\noindent
(a) $.1 \lsim x_F \lsim .3$

\noindent
The effective power is $\alpha \simeq .92 \pm
.01$ in this region \cite{cha}. The value of $\alpha$ for inclusive $\jpsi$ and
$\psi'$ production is found to be the same within errors. This is expected
since
at the high beam energies involved the $c\bar c$ pair remains compact
$(r_\perp \simeq 1/m_c \ll r_{\jpsi})$ inside the nucleus, with charmonium
formation occurring long after the pair has left the nucleus.

The $A$-dependence for $\Upsilon$ production is characterized by $\alpha
\simeq .97$. Since this is much closer to unity than in the case of charmonium,
the deviation of $\alpha$ from 1 is apparently due to a higher twist effect.
For
the Drell-Yan process of lepton pair production $\alpha \simeq 1$ \cite{dy}, as
it is for open charm
$(D)$ production \cite{e769,e789}. This suggests that (elastic) scattering
of the
heavy quarks in the nucleus, which can increase their relative
momentum, may be a reason for the nuclear suppression. There is no such
scattering for leptons, and it is irrelevant for open heavy quark
production. The more compact the pair is (\ie, the heavier the quark
mass,) the harder must the secondary elastic scattering be to resolve
the pair. This would explain the higher twist nature of the effect.

\noindent
(b) $x_F \gsim .4$

\noindent
In this region, $\alpha$ is found to decrease with
$x_F$, with $\alpha(x_F\simeq .6) \simeq .8$ \cite{cha}. Such $x_F$ dependence
is inconsistent with QCD factorization of the cross section into a
product of the hard subprocess and beam and target structure functions
\cite{hv}.
Since factorization should hold at the leading twist level \cite{qs}, this
indicates that there are important higher twist effects also in this region of
$x_F$. The suppression of $\Upsilon$ production is again found to be
appreciably
less than for the $\jpsi$, which is consistent with this conclusion.

An early suggestion for the $x_F$-dependent suppression of charmonium
production was energy loss due to gluon radiation from secondary
interactions in the nucleus \cite{gm}. There is, however, a general limit
to such
radiation set by the uncertainty principle \cite{bh}. Only gluons whose
formation
time are commensurate with the nuclear radius $R_A$ can be emitted inside
the nucleus. This imposes a limit on the momentum carried by such gluons,
$x_g \lsim <p_\perp^2> R_A/2E$.
At beam energies $E$ of several hundred GeV this implies negligible
energy loss, if one assumes an average hardness $<p_\perp^2> \simeq
0.1\,\gev^2$ for the secondary scattering in the nucleus. It has recently
been proposed \cite{jr} that the secondary nuclear scattering could be much
harder, which if true would be very interesting.

Alternatively, the suppression in the large $x_F$ region may be due to
scattering not from the heavy quarks, but from the low $x$ `stopped' light
quarks
which transferred their momentum to the heavy pair \cite{bhmt}. In the limit
where $m_c^2 \propto 1/(1-x_F)$, the light quarks are coherent with the heavy
pair. Hence scattering from the light quarks can put the heavy quarks on their
mass shell. The scattering cross section from the light quarks is large,
implying
a dominance of surface scattering on the nuclear target, \ie, $\alpha \simeq
2/3$. A mechanism of this type seems in any case to be required to explain
why the
$\jpsi$ becomes longitudinally polarized in $\pi N$ interactions as $x_F \to 1$
\cite{cip}. It is likely that this is due to helicity conservation from the
pion projectile to the leading $\jpsi$, which requires the full pion Fock
state to
interact coherently \cite{bb}.

In inelastic $\jpsi$ photoproduction the nuclear target dependence has been
measured as $\alpha = .99 \pm .04$ (for $p_\perp^2>1\ \gev^2$) by E691
\cite{e691}, and as $\alpha = 1.05 \pm .03$ (for $x_F \leq 0.85,\
p_\perp^2>0.4\
\gev^2$) by NMC \cite{nmc}. The photoproduced $\jpsi$'s are dominantly
produced at
large $x_F$. The absence of a nuclear suppression may be partly due to the cut
in $p_\perp$, which tends to enhance nuclear rescattering. Another reason may
be
the simpler Fock state structure of the photon, compared to hadrons (\cf\ the
remarks above on coherent scattering of hadron projectiles).

\noindent
(c) $x_F \lsim 0$

\noindent
The value of $\alpha$ measured in $pA \to \jpsi+X$ is found to decrease also as
one approches the nuclear fragmentation region \cite{cha}. The suppression
observed for $\Upsilon$ production is similar to that for charmonium. A
tentative
explanation for this is comover interactions -- partons moving with similar
velocities as the heavy quarks interact strongly with them, and may suppress
quarkonium formation \cite{com}. Since there are more comovers in the
fragmentation region of heavy nuclei, the comover effect should increase
with $A$.

At a proton beam energy of 800 GeV a $\jpsi$ with $x_F \simeq 0$ has an
energy of about 60 GeV in the nuclear rest frame. The energy $E_{com}$ of a
light
comover of similar velocity is reduced by a factor $<p_\perp>/M_{\jpsi}$, thus
$E_{com}\simeq 6$ GeV. Such a comover would be a `wee' parton in the
nuclear wave function. A better understanding of the comover effects may be
provided by experiments where a heavy ion beam is scattered on a hydrogen
target,
so that the nuclear fragmentation region is experimentally more accessible.

It should be clear already from the above brief review (from which several
other
outstanding puzzles of quarkonium physics were omitted due to the constraints
of
space) that charmonium production is a rich field of study, and that
nuclear targets can give us valuable information on the scattering dynamics. At
Elfe energies, one will be able to study the effects of charmonium formation
inside the nucleus. The experimental observation of charmonia with low
laboratory energies is thus an important challenge.

\vspace{.5cm}
\noindent
{\large\bf Acknowledgements}

I have benefitted from collaborations and discussions with, in particular, S.
J.
Brodsky, W.-K. Tang, R. Vogt and M. V\"anttinen. I am grateful to the
organizers
of this workshop, and in particular to Steven Bass, for their kind invitation.

\vfill\break

\end{document}